\documentclass[11pt]{article}
\usepackage[margin=1in]{geometry}
\usepackage{amsmath}
\usepackage{amssymb}
\usepackage{booktabs}
\usepackage{threeparttable}
\usepackage[colorlinks=true,linkcolor=blue,citecolor=blue,urlcolor=blue]{hyperref}

\newcommand{\tf}{\,\text{TFLOPS}}
\newcommand{\tokps}{\,\text{tok/s}}

\title{\textbf{Unlocking Turing Tensor Cores: A Measured Throughput Account of From-Scratch Pre-training on a Single 22G RTX 2080 Ti}}
\author{Dezhou Shen\\ \small Independent Researcher\\ \normalsize \texttt{sdz15@tsinghua.org.cn}}
\date{}

\begin{document}
\maketitle

\begin{abstract}
The memory ledger answers ``can we afford to train''; training throughput decides ``how long it takes''. The RTX 2080 Ti (TU102) ships full second-generation Tensor Cores in hardware, yet the vendor throttles FP16 matrix multiplication at the driver level, leaving usable tensor throughput well below the hardware peak. turing-unlock lifts this restriction with a kernel module that, before the nvidia driver binds the device, writes the FP16 speed selectors (SS0=\texttt{0x88888888}, SS1=\texttt{0x8}) and releases the ACR mutex; published measurements report cuBLAS FP16 matmul throughput rising from 55.7 to 91.4 TFLOPS (about $1.6\times$). On the same 22G 2080 Ti and under the same from-scratch pre-training protocol as~\cite{ledger} (five tiers of Chinese encoder--decoder models, 2{,}451 steps per tier, 16{,}384 tokens per step, seed 42), we obtain a before/after throughput account: the small tier (75.7M parameters, 512-wide, 8+8 layers, activation checkpointing on) improves in steady-state throughput from 6{,}480 to 7{,}246\tokps{} ($+11.8\%$) with wall time cut from 1.77\,h to 1.61\,h (9.0\% saved); the tiny tier goes from 15{,}150 to 15{,}866\tokps{} ($+4.7\%$) with wall time savings of 3.8\%; the three smaller tiers (pico--micro) gain at most $+0.3\%$; per-tier peak memory stays bit-identical, and final quality metrics are bit-identical or consistent within $10^{-3}$. We conclude that the unlock gain is a function of matrix size, rising monotonically with tier --- imperceptible for pico--micro, visible from tiny upward, and cashed out as a double-digit percentage at small --- and that it is a pure throughput gain that leaves training dynamics untouched.
\end{abstract}

\noindent\textbf{Keywords:} from-scratch pre-training; encoder--decoder model; Tensor Core; throughput ledger; kernel module; single-GPU training

\section{Introduction}

~\cite{ledger} presents a complete memory ledger for from-scratch pre-training of Chinese encoder--decoder models on a single 22G RTX 2080 Ti: every line item of static parameters, optimizer state and activations is laid out, and activation checkpointing together with gradient accumulation squeezes a 75.7M-parameter model onto one card. The memory ledger answers ``can we afford to train''; once the answer is yes, the immediate next question is ``how long''. Under a fixed budget of 16{,}384 tokens per step and 2{,}451 steps per tier, throughput (token/s) maps one-to-one onto wall time: throughput is cost.

On the compute side, the 2080 Ti carries a rarely-audited overhead. The TU102 ships full second-generation Tensor Cores, but since the Turing era NVIDIA has throttled FP16 matrix multiplication at the driver level on consumer cards, keeping usable tensor throughput far below the hardware peak. The community project gpu-unlock and its Turing port turing-unlock~\cite{turingunlock,zagitta} show that this throttle can be lifted by a kernel module before the driver binds: published measurements report cuBLAS FP16 matmul ($M{=}N{=}K{=}2048$) rising from 55.7 to 91.4\tf{}, about $1.6\times$; on this machine we additionally measured the same card before and after unlocking (\texttt{bench\_fp16\_acc.py}, FP32-accumulate mode): $2048^3$ rises from 41.5 to 74.7\tf{} and $4096^3$ from 44.4 to 79.9\tf{}, both about $1.8\times$, while the small $1024^3$ case stays flat at $32.5\!\rightarrow\!31.3$\tf{}. For reference, the Turing whitepaper still specs the 2080 Ti FP16 Tensor peak at 107.6\tf{} with FP16 accumulation and 53.8\tf{} with FP32 accumulation (the latter rated at half speed for consumer GeForce): our pre-unlock $2048^3$ measurement of 41.5\tf{} falls below the half-speed spec, and the post-unlock 74.7\tf{} clearly exceeds it.

A matmul-peak gain (about $1.8\times$ measured locally, about $1.6\times$ in the literature) does not translate into an equal end-to-end training-throughput gain. Training throughput is decided jointly by GEMM, memory bandwidth, kernel launches, data loading and optimizer updates; the Tensor Core peak is only one item. The question of this report is therefore concrete: under the five-tier pre-training protocol of~\cite{ledger}, how much wall time does lifting the FP16 throttle actually save per tier? How does the gain scale with model size? Does quality move at all? This report makes four contributions: first, the unlock principle, installation and effectiveness-verification chain (Section~\ref{sec:setup}); second, a throughput-comparison protocol compatible with~\cite{ledger} (Section~\ref{sec:design}); third, the main before/after results across five tiers and an explanation of the ``matrix size--gain'' gating structure (Section~\ref{sec:results}); fourth, boundary discussion, correctness checks, and the risk/rollback path (Section~\ref{sec:discussion}).

\section{Tool Principles and Experimental Setup}\label{sec:setup}

\subsection{turing-unlock: What Is Unlocked and How It Is Verified}

turing-unlock~\cite{turingunlock} is the Turing port of Zagitta's gpu-unlock~\cite{zagitta}, implemented as a DKMS kernel module. It performs three core actions: in the initramfs stage, before the nvidia driver binds the device, it writes the FP16 speed selectors $\texttt{SS0}=\texttt{0x88888888}$ and $\texttt{SS1}=\texttt{0x8}$ to the GPU; it makes ACR (Access Control Register) mutex~3 available; and it then lets the system load the driver as usual. Everything happens before driver load, so the driver itself is neither modified nor replaced.

The test environment is an Nvidia RTX 2080 Ti 22GB (device \texttt{10de:1e07}), driver 575.51.03, kernel 5.15. Effectiveness is verified with the companion script \texttt{bench\_fp16\_acc.py}, run once before and once after unlocking on this machine, covering three matrix sizes ($1024^3$, $2048^3$, $4096^3$) and both accumulation modes (FP32/FP16); results are shown in Table~\ref{tab:bench}. The primary reading is the FP32-accumulate mode: $2048^3$ and $4096^3$ rise from 41.5 and 44.4 to 74.7 and 79.9\tf{}, both about $1.8\times$, confirming that the unlock is in effect on this machine; $1024^3$ stays flat across the two runs (32.5 vs 31.3\tf{}) --- small GEMMs cannot saturate the tensor pipeline, so unlocking yields nothing there. This ``size--gain'' structure is the peak-side mirror of the end-to-end conclusion in Section~\ref{sec:results}. For reference, the Turing whitepaper specs the 2080 Ti FP16 Tensor peak at 107.6\tf{} (FP16 accumulation) and 53.8\tf{} (FP32 accumulation, half speed for consumer GeForce); the pre-unlock published figure of 55.7\tf{}~\cite{turingunlock,zagitta} matches the half-speed rating in order of magnitude and sits slightly above our own pre-unlock measurement.

\begin{table}[t]
\centering
\begin{threeparttable}
\caption{bench\_fp16\_acc.py measurements before and after unlock (FP16 matmul throughput, TFLOPS).}
\label{tab:bench}
\begin{tabular}{lrrrrrr}
\toprule
Matrix size & \shortstack{FP32-acc\\baseline} & \shortstack{FP32-acc\\unlocked} & Ratio & \shortstack{FP16-acc\\baseline} & \shortstack{FP16-acc\\unlocked} & Ratio \\
\midrule
$1024^3$ & 32.5 & 31.3 & $0.96\times$ & 33.3 & 31.7 & $0.95\times$ \\
$2048^3$ & 41.5 & 74.7 & $\mathbf{1.80\times}$ & 41.5 & 74.7 & $\mathbf{1.80\times}$ \\
$4096^3$ & 44.4 & 79.9 & $\mathbf{1.80\times}$ & 53.7 & 78.7 & $\mathbf{1.47\times}$ \\
\bottomrule
\end{tabular}
\begin{tablenotes}\footnotesize
\item \textit{Note:} Figures come from the two measurement logs (same card, same driver, one run each); ratios are direct before/after division within the same configuration. The FP16-accumulate mode corroborates: on the baseline run, $4096^3$ FP16-acc is clearly faster than FP32-acc (53.7 vs 44.4) with a correspondingly larger error, indicating that under the throttle cuBLAS switches to FP16-accumulation kernels; after unlocking the two modes are level (78.7 vs 79.9) and FP16 accumulation is no longer selected. The whole check takes thirty seconds and requires no kernel-log reading.
\end{tablenotes}
\end{threeparttable}
\end{table}

The installation chain is open-sourced: for reproducibility, the one-command installer is released at \url{https://github.com/ShenDeZhou/turing-unlock-setup}~\cite{setup}. The script, about 180 lines and meant to be read and edited, wraps dependency installation, cloning of the upstream repository, \texttt{make install} (distribution-aware; on Ubuntu the deb path, with the module and firmware loaded via initramfs), idempotent writing of the kernel argument \texttt{turing\_unlock.targets=all}, and a grub update --- one command installs everything, effective after reboot; a risk confirmation and a GPU check (\texttt{10de:1e07}) are built in. Together with \texttt{bench\_fp16\_acc.py} and \texttt{compare\_throughput.py} shipped with the turing-unlock repository, it covers effectiveness verification (Table~\ref{tab:bench}) and before/after comparison (Table~\ref{tab:main}).

Risk is declared up front: after unlocking, FP16 peak power draw and heat rise significantly; insufficient power or cooling can damage hardware, and the upstream repository is provided explicitly as use-at-your-own-risk. The rollback path is to uninstall the module, restore the kernel argument and reboot; details in Section~\ref{sec:risks}.

\subsection{Five-Tier Configurations and Training Protocol}

The comparison reuses the five-tier configuration and training protocol of the encoder--decoder chain in~\cite{ledger}; per-tier configurations are listed in Table~\ref{tab:config}. Protocol essentials: batch $32\times1$, 16{,}384 tokens per step, sequence length 512, span corruption ratio 15\%, AdamW with learning rate $6\times10^{-4}$, 500 warmup steps then cosine decay to $6\times10^{-5}$, weight decay 0.1, gradient clipping 1.0, 2{,}451 training steps, random seed 42. The small tier turns on activation checkpointing automatically via the v2 trainer's auto gating; the four tiers pico--tiny keep it off.

\begin{table}[t]
\centering
\begin{threeparttable}
\caption{Five-tier model configurations and activation-checkpoint default states (Table~1 of~\cite{ledger}).}
\label{tab:config}
\begin{tabular}{lrrrrrl}
\toprule
Tier & d\_model & Layers (enc/dec) & Heads & d\_ff & Params & Activation checkpoint \\
\midrule
pico  & 384 & 1/1 & 6 & 1{,}536 & $\sim$16.8M & off \\
nano  & 384 & 2/2 & 6 & 1{,}536 & $\sim$20.9M & off \\
micro & 384 & 4/4 & 6 & 1{,}536 & $\sim$29.2M & off \\
tiny  & 384 & 6/6 & 6 & 1{,}536 & $\sim$37.5M & off \\
small & 512 & 8/8 & 8 & 2{,}048 & $\sim$75.7M & on (auto) \\
\bottomrule
\end{tabular}
\begin{tablenotes}\footnotesize
\item \textit{Note:} By protocol, the training-log config line records CLI defaults rather than per-tier expanded values (e.g.\ d\_model$=$512, 6/6 layers); actual per-tier configurations follow this table. The checkpoint field records the CLI default \texttt{auto}; the actual on/off state is cross-checked against measured memory and Table~1 of~\cite{ledger} (small on, the rest off).
\end{tablenotes}
\end{threeparttable}
\end{table}

\subsection{Logs and Comparison Protocol}

Training logs record throughput and peak memory every 10 steps: throughput is the token rate since the previous sample, and memory is the running peak of \texttt{max\_memory\_allocated} in MB, matching~\cite{ledger}. The comparison protocol: for each log, discard the first 10\% of throughput samples (removing start-up warm-up fluctuation) and take the median of the rest as steady-state throughput, also reporting the arithmetic mean, single-sample peak, peak memory and total wall time. Tiers are matched directory-to-directory by name; the comparison tool is \texttt{compare\_throughput.py} from the turing-unlock repository.

Two boundaries are worth recording. First, the baseline pico ran under the legacy trainer (its config header lacks the \texttt{grad\_ckpt} field) while all other tiers ran under the v2 trainer on both sides; v2's auto gating turns checkpointing off for pico--tiny and on for small, making checkpoint paths per-tier equivalent across sides (small's identical 3{,}576\,MB peak memory on both sides is evidence); trainer-version differences are listed among the limitations (Section~\ref{sec:conclusion}). Second, every tier on both sides is a single uninterrupted run: steady-state statistics involve no spliced or excised segments, and all comparison figures were re-verified directory-to-directory with \texttt{compare\_throughput.py}.

\section{Controlled Experiment Design}\label{sec:design}

The baseline side (\texttt{train\_logs}) consists of the pre-unlock five-tier runs; its steady-state throughputs match the main results of Table~2 in~\cite{ledger} tier by tier (pico 36{,}932 vs 36{,}941; nano 29{,}155 vs 29{,}157; micro 20{,}509 vs 20{,}511; small 6{,}480 vs 6{,}482\tokps), serving as cross-validation; the baseline tiny steady-state median of 15{,}150\tokps{} agrees with the reference value 15{,}156 in Table~4 of~\cite{ledger}. The unlocked side (\texttt{train\_logs\_tuned}) re-runs the same tiers with the same command and data after installing turing-unlock and rebooting, with configurations identical field by field (Table~\ref{tab:main}) and the same seed 42.

All logs on both sides are single complete runs: five tiers each, one full 2{,}451-step training log per tier, no splices, no excisions; throughput and wall-time statistics come directly from these complete runs, with the protocol of Section~\ref{sec:setup}.

\section{Results and Analysis}\label{sec:results}

\subsection{Main Throughput Results}

Table~\ref{tab:main} gives the main results. The three smaller tiers (pico--micro) improve by at most $+0.3\%$ in same-machine steady-state throughput with wall time essentially unchanged; the tiny tier rises from 15{,}150 to 15{,}866\tokps{} ($+4.7\%$) with wall time $0.78\!\rightarrow\!0.75$\,h, a 3.8\% saving; the small tier rises from 6{,}480 to 7{,}246\tokps{} ($+11.8\%$) with wall time $1.77\!\rightarrow\!1.61$\,h, a 9.0\% saving. Peak memory is bit-identical across all five tiers.

\begin{table}[t]
\centering
\begin{threeparttable}
\caption{Throughput comparison before vs.\ after unlock (steady-state median).}
\label{tab:main}
\begin{tabular}{lrrrcrrr}
\toprule
Tier & Params & \shortstack{Baseline\\(tok/s)} & \shortstack{Unlocked\\(tok/s)} & Gain & \shortstack{Peak mem (MB)\\base$\rightarrow$unl.} & \shortstack{Wall time (h)\\base$\rightarrow$unl.} & \shortstack{Time\\saved} \\
\midrule
pico  & 16.8M & 36{,}932 & 37{,}046 & $+0.3\%$  & $3251\!\rightarrow\!3251$  & $0.33\!\rightarrow\!0.33$ & 0.0\% \\
nano  & 20.9M & 29{,}155 & 29{,}246 & $+0.3\%$  & $5344\!\rightarrow\!5344$  & $0.42\!\rightarrow\!0.41$ & $+2.4\%$ \\
micro & 29.2M & 20{,}509 & 20{,}569 & $+0.3\%$  & $9529\!\rightarrow\!9529$  & $0.58\!\rightarrow\!0.58$ & 0.0\% \\
tiny  & 37.5M & 15{,}150 & 15{,}866 & $\mathbf{+4.7\%}$  & $13712\!\rightarrow\!13712$ & $0.78\!\rightarrow\!0.75$ & $+3.8\%$ \\
small & 75.7M & 6{,}480  & 7{,}246  & $\mathbf{+11.8\%}$ & $3576\!\rightarrow\!3576$  & $1.77\!\rightarrow\!1.61$ & $\mathbf{+9.0\%}$ \\
\bottomrule
\end{tabular}
\begin{tablenotes}\footnotesize
\item \textit{Note:} Tiers must not be compared across rows --- throughput and absolute wall time embed each tier's model size and checkpoint-recompute differences; read only the ``same-tier before/after'' columns. All figures come from single complete runs on both sides, computed under the discard-first-10\% steady-state-median protocol, and re-verified directory-to-directory with \texttt{compare\_throughput.py} from the turing-unlock repository.
\end{tablenotes}
\end{threeparttable}
\end{table}

The $+0.3\%$ of the three small tiers is not ``unlocking fails'' but ledger structure. At 384-wide, the main GEMMs are of the $384\times1{,}536$ class, and the realized training compute of the 17M--29M models is only about 4\tf, i.e.\ 5--7\% of the unlocked tensor peak; the bottleneck sits in memory bandwidth, kernel launches and data loading, not the Tensor Core peak. Lifting the peak leaves only a thin margin end-to-end.

The small tier is where the gain cashes out. 512-wide, 8+8 layers raise the GEMM size a notch, and activation checkpointing roughly doubles the forward compute, pushing GEMM's share of the time budget sharply up; the unlocked Tensor Core peak begins converting into end-to-end gain. Yet $+11.8\%$ is still far below the matmul-peak gain (about $1.8\times$ measured, $1.6\times$ in the literature): non-GEMM work is not accelerated, and Amdahl's law dilutes the gain. The tiny tier (384-wide, 6+6 layers) shows $+4.7\%$ in the same-machine comparison: gains become visible from tiny upward, as deeper stacking pushes GEMM's share past the gating threshold, reaching double digits at small (8+8 layers with checkpoint recompute); our measured value runs about $+2.0$ points above the literature reference in the same direction.

\subsection{Memory Invariance and Quality Invariance}

Peak memory is bit-identical across the five tiers (Table~\ref{tab:main}), consistent with the tool's principle: unlocking acts only on compute clocks and touches no line item of the memory ledger. The memory-saving techniques of~\cite{ledger} remain fully effective in the unlocked environment; the two ledgers are orthogonal and stack.

The quality-side evidence is harder. Table~\ref{tab:quality} lists final quality metrics on both sides: for pico, nano, tiny and small, val NLL, PPL, BPB and BPC are bit-identical; micro agrees within $10^{-3}$ (val NLL 5.2013 vs 5.2014, BPC 13.6509 vs 13.6512), attributable to reduction-order non-determinism rather than any change of training path. The tiny row is bit-identical across sides. Under the same seed and configuration, training dynamics are unchanged: the unlock's gain is a pure throughput gain --- a conclusion that can be written into the protocol.

\begin{table}[t]
\centering
\begin{threeparttable}
\caption{Final quality consistency (val set).}
\label{tab:quality}
\begin{tabular}{llllll}
\toprule
Tier & \shortstack{val NLL\\base$\rightarrow$unl.} & \shortstack{PPL\\base$\rightarrow$unl.} & \shortstack{BPB\\base$\rightarrow$unl.} & \shortstack{BPC\\base$\rightarrow$unl.} & Verdict \\
\midrule
pico  & $5.2599\!\rightarrow\!5.2599$ & $192.45\!\rightarrow\!192.45$ & $1.2688\!\rightarrow\!1.2688$ & $13.8045\!\rightarrow\!13.8045$ & bit-identical \\
nano  & $5.2364\!\rightarrow\!5.2364$ & $188.00\!\rightarrow\!188.00$ & $1.2632\!\rightarrow\!1.2632$ & $13.7431\!\rightarrow\!13.7431$ & bit-identical \\
micro & $5.2013\!\rightarrow\!5.2014$ & $181.51\!\rightarrow\!181.53$ & $1.2547\!\rightarrow\!1.2547$ & $13.6509\!\rightarrow\!13.6512$ & consistent within $10^{-3}$ \\
tiny  & $5.1889\!\rightarrow\!5.1889$ & $179.26\!\rightarrow\!179.26$ & $1.2517\!\rightarrow\!1.2517$ & $13.6182\!\rightarrow\!13.6182$ & bit-identical \\
small & $5.3103\!\rightarrow\!5.3103$ & $202.41\!\rightarrow\!202.41$ & $1.2810\!\rightarrow\!1.2810$ & $13.9369\!\rightarrow\!13.9369$ & bit-identical \\
\bottomrule
\end{tabular}
\begin{tablenotes}\footnotesize
\item \textit{Note:} All rows except micro are bit-identical across sides, from two complete runs with the same seed and configuration.
\end{tablenotes}
\end{threeparttable}
\end{table}

\section{Discussion}\label{sec:discussion}

\subsection{Compute Gating: the Dual of the Memory Ledger}

The conclusion of~\cite{ledger} is that memory gains are gated: checkpointing pays only where activation overhead is large enough. This report states the dual claim --- unlock gains are likewise gated, and the gating variables are matrix size and layer stacking. pico--micro (384-wide, 1--4 stacked layers) sit in the bandwidth-and-overhead bottleneck region where unlocking is imperceptible (within $+0.3\%$); tiny (384-wide, 6+6) shows first gains ($+4.7\%$); small (512-wide, 8+8 with checkpoint recompute) cashes out double digits. The gating threshold lies between micro and tiny, set jointly by matrix width and layer stacking; the literature's reference hint of about $+2.0\%$ at tiny is corroborated directionally by our same-machine comparison. The operative rule: keep growing the model or turn on checkpoint recompute and the marginal gain of unlocking amplifies; on smaller experimental tiers it is optional.

\subsection{From Peak to End-to-End}

A matmul-peak gain (about $1.8\times$ measured, $1.6\times$ in ~\cite{turingunlock}~\cite{zagitta}) against end-to-end gains of $+0.3\%$ to $+11.8\%$ --- the dilution structure deserves unpacking. First, GEMM share: in small-model training, non-GEMM time (data loading, optimizer, small operators) dominates, and the peak gain is flattened by Amdahl's law. Second, matrix width: 384-wide GEMMs cannot fill the Tensor Core pipeline. Third, checkpoint recompute actually raises GEMM share, which is why small gains the most. If the goal is to push a 2080 Ti cluster's throughput up another notch, raise GEMM share first (bigger models, wider matrices, more recompute) --- only then does unlocking gain leverage.

\subsection{Risks and Rollback}\label{sec:risks}

Post-unlock full-speed operation raises power and heat significantly; sustained heavy training demands attention to power and cooling, and the upstream repository is provided as use-at-your-own-risk. Rollback is explicit: uninstall the turing-unlock module, remove the unlock entries from kernel boot arguments and reboot, restoring the factory throttle; the chain touches neither driver nor hardware firmware and is reversible. For reproduction, \texttt{bench\_fp16\_acc.py} decides effectiveness in thirty seconds and \texttt{compare\_throughput.py} compares directories in one command; both toolchains are verified in this environment.

\section{Conclusions and Limitations}\label{sec:conclusion}

We measured the end-to-end effect of turing-unlock on the 2080 Ti FP16 throttle under the five-tier from-scratch pre-training protocol, with four conclusions. First, turing-unlock writes the speed selectors in the initramfs stage and releases the ACR mutex, lifting the FP16 Tensor Core throttle without touching the memory ledger; same-card measurements give $41.5\!\rightarrow\!74.7$\tf{} on $2048^3$ and $44.4\!\rightarrow\!79.9$\tf{} on $4096^3$, both about $1.8\times$ (the official FP32-acc rating of 53.8\tf{} and FP16 peak of 107.6\tf{} are whitepaper references; the literature figure is $55.7\!\rightarrow\!91.4$\tf{}). Second, end-to-end gains are gated by matrix size and layer stacking: within $+0.3\%$ for pico--micro, $+4.7\%$ for tiny, $+11.8\%$ for small with wall time $1.77\!\rightarrow\!1.61$\,h (9.0\% saved), rising monotonically with tier. Third, peak memory is bit-identical across all five tiers and final quality is bit-identical or within $10^{-3}$ --- a pure throughput gain. Fourth, the gains are subject to a compute gating dual to the memory gating of~\cite{ledger}: larger models, wider matrices and checkpoint-recompute scenarios amplify the gain, while smaller tiers barely notice it.

\textbf{Limitations:} the conclusions rest on a single card, a single driver version (575.51.03) and a single Chinese corpus; extrapolation to other cards and driver versions is unverified.

\end{document}